# Three +1 Faces of Invariance

Moses Fayngold
*Department of Physics, New Jersey Institute of Technology, Newark, NJ 07102*

A careful look at an allegedly well-known century-old concept reveals interesting aspects in it that have generally avoided recognition in literature. There are four different kinds of physical observables known or proclaimed as relativistic invariants under space-time rotations. Only observables in the first three categories are authentic invariants, whereas the single "invariant" – proper length – in the fourth category is actually not an invariant. The proper length has little if anything to do with *proper distance* which is a true invariant. On the other hand, proper distance, proper time, and rest mass have more in common than usually recognized, and particularly, mass – time analogy opens another view of the twin paradox.

## *Introduction*
## *Use and Abuse of the concept of Invariance*

The importance of invariant characteristics of physical systems or processes cannot be overstated. The role invariants play in the description of the world is best captured by E. Taylor and J. A. Wheeler [1] in the statement:

"*In relativity, invariants are diamonds. Do not throw away diamonds*!"

Invariants are an indispensible working tool of every professional physicist and an unavoidable topic in almost any Physics textbook.

With all that, the concept of invariance and invariant characteristics turns out to be more subtle than usually perceived. There are some important aspects of invariance which, to my knowledge, have not been explicitly stated in literature. In particular, what is usually referred to as just "invariants" actually fall into a few quite different categories. Ignoring this fact, as shown below, leads to confusion, and frequently causes misleading or downright wrong statements. This paper outlines four distinct kinds of invariants – four faces of Lorentz-invariance – and shows that one of them is just a misconstruction reflecting a widely spread misconception, which is still being disseminated among generations of Physics students.

Sorting out the known invariant characteristics of physical systems by their properties also helps elucidate very close analogies between some of them. Specifically, the analogy between mass and time – the two observables describing totally different aspects of reality – will enable us to see the twin paradox from a different angle.

## I. *Face* 1

The first category can be called the operational invariance. Most physical characteristics of a system or process (e.g., mass, size, time, etc.) can be measured using an executable *experimental* procedure which can be performed in any inertial reference frame (RF) and/or for any state of motion of the studied system. If such a procedure finds a certain quantity numerically the same in any RF, then this quantity is an *operational invariant*.

The simplest example is a number $N$ of stable particles in an isolated system, which is the same for all observers[1].

Another invariant in this category is the electric charge $Q$. Its independence from velocity is evident in many known experiments. As an example, consider the electron-positron annihilation $e + e^{+} \rightarrow \gamma + \gamma$ under two different conditions: first, in a $(e, e^{+})$ "atom" when the velocities of both particles are negligible; and second, in a high-energy collision with only one particle ("target") stationary and the other one moving with an ultra-relativistic speed. In both cases the net charge of the system after annihilation is zero, which (assuming the net charge is conserved) shows that the charge of the bombarding particle is the same as in its stationary state. By reciprocity this also means its independence from the observer's state of motion. This puts the charge into the category of operational invariants.

## II. *Face* 2

Next we consider the most famous of all invariants – the speed of light in vacuum. It has a rather subtle distinction from the operational invariants. On the one hand, it seems to fall into their category I through a procedure which is the same in all RF (e.g., the Michelson experiment) and gives the same result in all of them. But on the other hand, the subject of direct measurement (a free photon with a definite momentum) does not have a rest frame. And it is not a quantity like charge $Q$ or occupancy $N$ whose independence from speed is established beyond any doubts. It is *the photon speed itself* that remains the same in any RF and for all photons in one RF regardless of their momentum. Moreover, it turns out to be the same for *all* objects of nature, if we, instead of 3-velocity $\mathbf{v}$, measure their 4-velocity $u$ [3, 4] defined as

$$u^j \equiv c\,\frac{dx^j}{ds} \quad , \quad j = 0, 1, 2, 3 \tag{2.1}$$

---

[1] We assume "regular" particles – photons and/or tardyons; for hypothetical tachyons, the situation may be different [2].

Here $ds$ is the norm of an incremental 4-displacement along the world line of the corresponding object (for the reasons discussed in [5], a more meaningful definition would be using $|ds|$ instead of $ds$). The norm of 4-velocity of *any* object turns out to be *universal constant* equal to the speed of light – unlike the operational invariants, it satisfies the identity:

$$u_j u^j = u_0^2 - \mathbf{u}^2 \equiv c^2 \tag{2.2}$$

Here subscripts and superscripts on the left stand respectively for the co- and contravariant components of a 4-vector, and Einstein's summation rule is used. In terms of 4-velocity, *all* objects of nature flow through space-time with the speed of light, and the only difference between a photon, a rocket, and a stationary rock in this regard is in the tilt of their respective world lines. Thus, the speed of light is more than just an invariant – it is a universal constant, the same for any object. Therefore it can be called an absolute or *universal invariant*, which, unlike $N$ or $Q$, is a single-valued quantity.

## III. ***Face* 3**

The invariants in the third category are the rest mass, proper time, and proper distance.

III $a$: **Rest mass**. Traditionally (before 1905) mass was defined as the ratio $f/a$ of a net external force $\mathbf{f}$ to resulting acceleration $\mathbf{a}$. Measurements at $v \ll c$ did not show any noticeable change of mass in a moving object. In this respect, mass was thought to be in one company with electric charge.

Relativity changed all this. Both – the theory and new experiments – showed that mass as a measure of body's inertness is velocity-dependent. Moreover, this dependence is different for different angles between $\mathbf{f}$ and $\mathbf{v}$. Measuring mass as the ratio $f/a$ when $\mathbf{f} \perp \mathbf{v}$, that is, $m = f_\perp / a_\perp$, gives the well-known expression

$$m = m_0\,\gamma(v) \quad , \quad \gamma(v) \equiv \left(1 - \frac{v^2}{c^2}\right)^{-1/2} \tag{3.1}$$

For $\mathbf{f} \parallel \mathbf{v}$, that is, for the ratio $f_\parallel / a_\parallel$, measurements give $m = m_0\,\gamma^3(v)$. In both expressions $m_0$ is the rest mass of the object.

Mass (3.1) determined for $\mathbf{f} \perp \mathbf{v}$, is known as the *transverse mass* and is identical to the *relativistic mass* discussed below; mass determined at $\mathbf{f} \| \mathbf{v}$, is the longitudinal mass [3, 4].

For arbitrary individual orientations of **f** and **v**, the resulting acceleration is not even parallel to the force, so **f** cannot generally be represented as $\mathbf{f} = m\mathbf{a}$ with a scalar-valued coefficient $m$ even as a function $m(\mathbf{v})$ [5, 6]. Nevertheless, the relativistic generalization of the second law can be formulated not only in terms of Minkowski's 4-force, but also in terms of 3-vectors **f** and **a** for any orientation between **f** and **v** [3-5]. Such formulation involves "anisotropic mass" as a second-rank $3\times 3$ tensor, which can be linked to the spatial part of the $4\times 4$ energy-momentum tensor, and is especially convenient in relativistic mechanics of continuous mediums [7-10].

Here it is sufficient to restrict to case (3.1). It has been extensively studied for charged particles in a magnetic field, for which the condition $\mathbf{f} \perp \mathbf{v}$ is automatically satisfied. The experiments showed an increase of mass with velocity [11] even before the advent of the theory of relativity, and later [12] they confirmed Eq. (3.1) unambiguously. The results were applied to the development of synchrotron accelerators [13, 14].

As mentioned above, Eq. (3.1) also defines *relativistic mass*. Apart from and *independently* of (3.1), relativistic mass is defined as the ratio $p/v$, where **p** is particle's momentum. We thus have two different procedures yielding the same result, and accordingly *two equivalent definitions*: relativistic mass (3.1) defined either as $f_\perp / a_\perp$ or as $p/v$; it is a speed-dependent and therefore not invariant characteristic. This is an *experimental fact* that cannot be dismissed.

The expression (3.1) for relativistic mass is consistent with the general mass-energy equivalence

$$E = mc^2 \quad , \tag{3.2}$$

where the relativistic energy $E$ is not an invariant either, but is only the temporal component of 4-momentum.

Speed dependence of the relativistic mass was, perhaps, one of the factors that triggered the question of whether such characteristic should be considered at all. There is a fashionable trend to consider only the rest mass as a legitimate description of an object, denying the relativistic mass the status of a meaningful characteristic [15–17]. According to this view, there is only the rest mass $m_0$, and relation (3.2) can only be applied to the rest mass and rest energy, respectively, that is, we must restrict it to

$$E_0 = m_0 c^2 \tag{3.3}$$

(here and hereafter the subscript "0" stands for a quantity measured in the rest frame of an object). This "truncation" of Eq. (3.2) is merely a viewpoint which cannot invalidate the equation itself. Indeed, expressing $E_0$ in (3.3) in terms of $E$ from the Lorentz-transformation of 4-momentum,

$$E = \gamma(V)\left(E_0 \pm V p_0\right) \underset{p_0 = 0}{=} \gamma(V)\, E_0 \,, \qquad E_0 = \frac{E}{\gamma(V)} \tag{3.4}$$

one immediately recovers (3.2) with $m = \gamma(V)\, m_0$. As Richard Feynman explicitly emphasized in his "*Lectures on Physics*" [18], "*The total energy of a particle is its mass in motion times* $c^2$ ($E = mc^2$), *and when the body stops, its energy is its rest mass times* $c^2$ ($E_0 = m_0 c^2$)"

The universal nature of mass-energy relation was emphasized in many sources as exemplified in [19]: "*Physical manifestations of the aspects of matter corresponding to mass and energy, respectively, are different; but the quantitative characteristics of these aspects are universally proportional to one another. It is this universal proportionality that allows one to speak about the mass-energy equivalence*".

Summarizing this part, we can say that the rest mass of an object is a scalar-valued coefficient converting its 4-velocity *u* into 4-momentum $\mathcal{P}$, whereas its relativistic mass is a scalar-valued (but speed-dependent) coefficient converting its 3-velocity **v** into 3-momentum **p**.

Already the mere fact that we have to distinguish between just mass and the rest mass shows that the rest mass is *not* an operational invariant defined in Sec. I. The existence of at least one verifiable experiment recording velocity dependence of mass (or any other observable) automatically takes it out of domain I. Within domain I, *we do not talk about the* "*rest*" *charge and* "*relativistic*" *charge*. Unlike charge, the rest mass is (up to the factor *c*) the norm of a 4-momentum $\mathcal{P} = \left(E/c\ ,\ \mathbf{p}\right)$:

$$m_0^2\, c^2 = \frac{E^2}{c^2} - \mathbf{p^2} = p_j\, p^j\ , \quad j = 0, 1, 2, 3 \qquad (3.5)$$

This is used for determining $m_0$ when it cannot be measured directly in its rest frame (e.g., in high-energy physics). We measure instead *E* and *p*, and then calculate $m_0$ from (3.5). The measurements must be very accurate since in the ultra-relativistic case the computed value of $m_0$ comes out as a small difference between two large numbers.

It is argued sometimes that the norm (3.5) (divided by *c*) must be taken as the *general definition* of mass. One cannot dispute a definition, but one can dispute its consistence with other elements of reality. In the real world, an entity called mass manifests itself through its inertia which is measurable under various conditions, and a special but important subset of conditions (the ratio $f_\perp / a_\perp$) gives the result (3.1) called the relativistic mass. This result is consistent with both –the expression $\left(p/v\right)$ and general Eq. (3.2). But definition (3.5), on the other hand, determines the rest mass exclusively, rather than mass in general.

Thus, the rest mass is a relativistic invariant without being an *operational* invariant. But its value measured in its *rest frame* can also be *computed* as the norm of the particle's 4-momentum measured in an *arbitrary* RF. This kind of invariant can be called a *rotational invariant* or a *norm-invariant* since it is not affected by rotations in space-time.

This becomes self-evident if we use the "banned" relationship (3.2) to rewrite (3.5) as

$$m_0^2 = m^2 - \frac{\mathbf{p}^2}{c^2} \qquad (3.6)$$

Since $\mathbf{p}^2 = p^2$ and $p/m = v$, we can also write this as

$$m_0^2 = m^2 \left(1 - \frac{p^2}{m^2 c^2}\right) = \frac{m^2}{\gamma^2\,(v)} \qquad (3.7)$$

Eq-s (3.5)–(3.7) clearly show that $m_0$ (not $m$!) is the norm of a 4-vector. The quantities $m$ and $\mathbf{p}/c$ on the right of (3.6) are the "temporal" and "spatial" projections of this 4-vector in the energy-momentum (or mass-momentum) space. In this interpretation, the velocity dependence of relativistic mass is a *natural geometrical effect*, since different velocities correspond to different 4-rotations of a given RF with respect to the rest frame of the object, and accordingly to different values of the temporal projection of its 4-momentum. Thus, we have the energy-momentum relation (3.5) or mass-momentum relation (3.6), but both terms, albeit expressing different aspects of matter, are equally legitimate, and which one to use is a matter of taste but not the matter of principle, in view of the total mathematical equivalence of (3.5) and (3.6).

III *b*: **Proper time**. Consider a time-like 4-displacement between two events:

$$s_{\mathrm{AB}}^2 = c^2 t_{\mathrm{AB}}^2 - \mathbf{r}_{\mathrm{AB}}^2 > 0 \tag{3.8}$$

Here A and B label the respective end-points of the interval $s_{\mathrm{AB}}$; $t_{\mathrm{AB}} \equiv t_{\mathrm{B}} - t_{\mathrm{A}}$ is the time between the events in a given RF, and $\mathbf{r}_{\mathrm{AB}} = \mathbf{r}_{\mathrm{B}} - \mathbf{r}_{\mathrm{A}}$ is the spatial displacement between them. In the *proper* frame $K_0$, where both events happen at one place ($\mathbf{r}_{\mathrm{AB}} = 0$), A and B are on its temporal axis. Setting $\mathbf{r}_{\mathrm{AB}} = 0$ in (3.8) gives

$$\tau_{\mathrm{AB}} = \frac{s_{\mathrm{AB}}}{c} \tag{3.9}$$

Here $\tau_{\mathrm{AB}}$ is the time lapse between the events in the proper frame, which is the definition of *proper time* between the events. For $\mathbf{r}_{\mathrm{AB}} \neq 0$, we can use (3.9) to rewrite (3.8) in the form analogous to (3.6), (3.7):

$$\tau_{\mathrm{AB}}^2 = t_{\mathrm{AB}}^2 - \frac{\mathbf{r}_{\mathrm{AB}}^2}{c^2} = \frac{t_{\mathrm{AB}}^2}{\gamma^2(V)}\ ; \quad t_{\mathrm{AB}} = \tau_{\mathrm{AB}}\ \gamma(V)\ ; \quad \mathbf{V} \equiv \frac{\mathbf{r}_{\mathrm{AB}}}{t_{\mathrm{AB}}} \tag{3.10}$$

The right side of this expression describes the time dilation effect. As seen from (3.10), the possibility of writing $\tau_{\mathrm{AB}}$ as the norm of the corresponding time-like 4-displacement rests on this effect. Like the rest mass, the proper time can be measured *directly* in $K_0$, or *computed* as the norm of the corresponding 4-interval, which puts it into the category of norm-invariants.

The mathematical similarity between III *a* and III *b* exposes an important analogy between the rest mass of an object and the proper time of a process, which, in my opinion, deserves an additional sub-section below.

III *ab*: **Rest mass – proper time analogy**

This is an analogy between a system of non-interacting particles, on the one hand, and a succession of consecutive processes within a single moving object, on the other. The analogy is easily seen and yet has, to my knowledge, been generally overlooked in literature.

Start with the rest mass. For a *system* of moving non-interacting particles, its rest mass $M_0$ is the sum of the *relativistic* masses $m_j$ of the particles in the system's rest frame (*not* the sum of their rest masses $m_{0j}$) [3, 5, 6, 15, 20]:

$$M_0 = \sum_j m_j = \sum_j m_{0j}\,\gamma(v_j) \;\geq\; \sum_j m_{0j} = m_0 \qquad (3.11)$$

(the sum on the right would amount to the rest mass of the system if all its particles are at rest). Inequality (3.11) is frequently interpreted as non-conservation of the system's rest mass. Such interpretation is totally misleading since the term "conserved" means some property of an isolated system, which remains constant during its *time evolution*. "Non-conservation" of the rest mass would mean, by virtue of Eq. (3.3), non-conservation of system's rest energy.

The appropriate statement about (3.11) is that the rest mass is a *non-additive* characteristic of a system [5]. The geometry of this non-additivity is illustrated in Fig. 1.

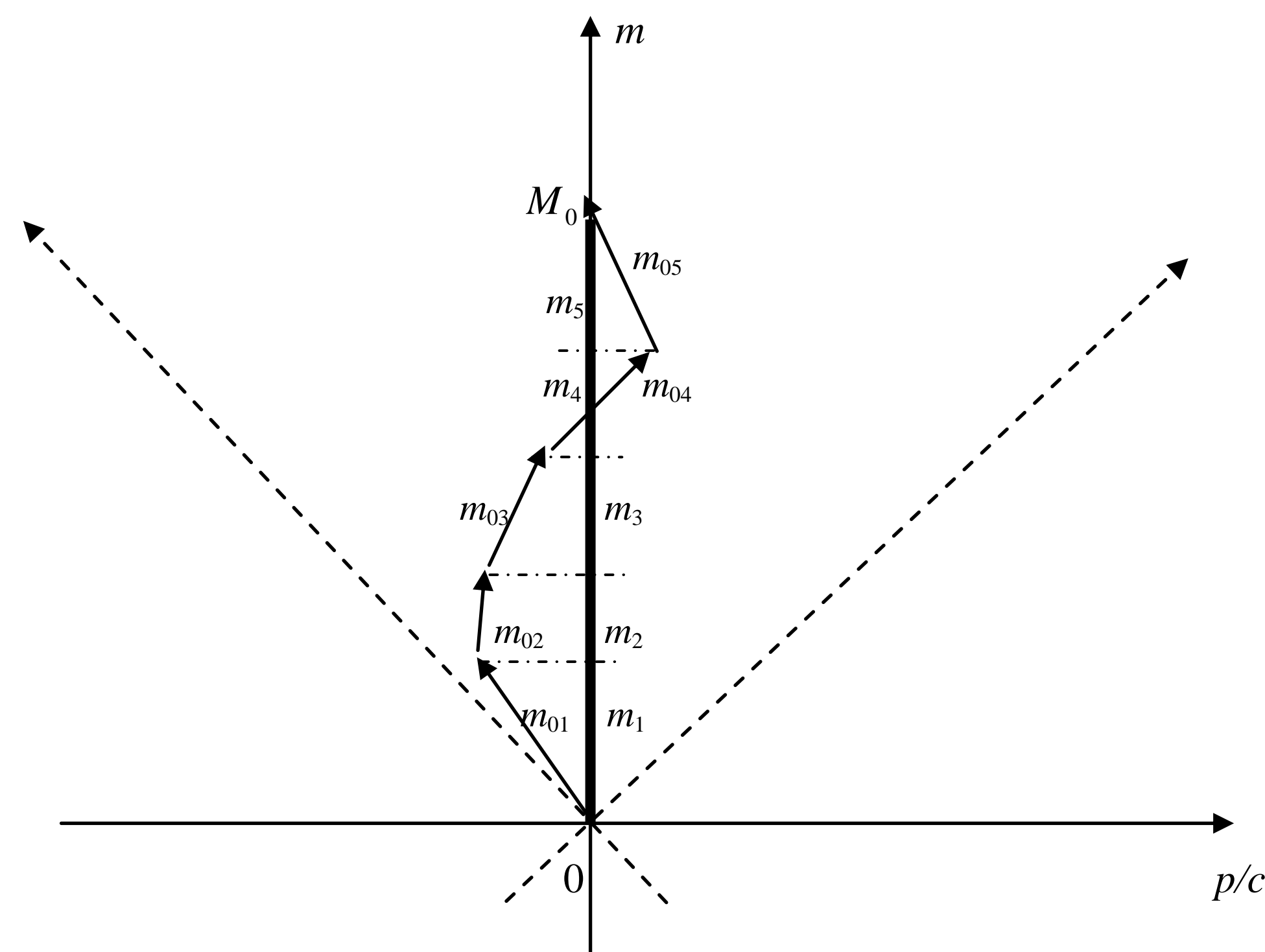

**Fig. 1**

Graph of the invariant (rest) mass $M_0$ of an object as the norm of its 4-momentum.
The vector is normalized to mass and shown in the object's rest frame. For a system of non-interacting parts, $M_0$ is the *geometric* (not algebraic!) sum (represented by segment $0M_0$) of their individual rest masses $m_{0j}$, which equals the algebraic sum of their relativistic masses $m_j$ (projections onto the *m*-axis). Vectors representing individual invariant masses $m_{0j}$ are generally not parallel to the *m*-axis because these masses are moving in the object's rest frame. The dashed lines represent photons' trajectories in the momentum space (or, which is the same, their zero rest masses).

Consider now a single moving object T (equipped with a clock), starting from a certain point in space (event A) and later returning to the same point (event B). If we choose A as the origin of the corresponding frame $K_0$, then both A and B lie on the time axis of this frame (Fig. 2).

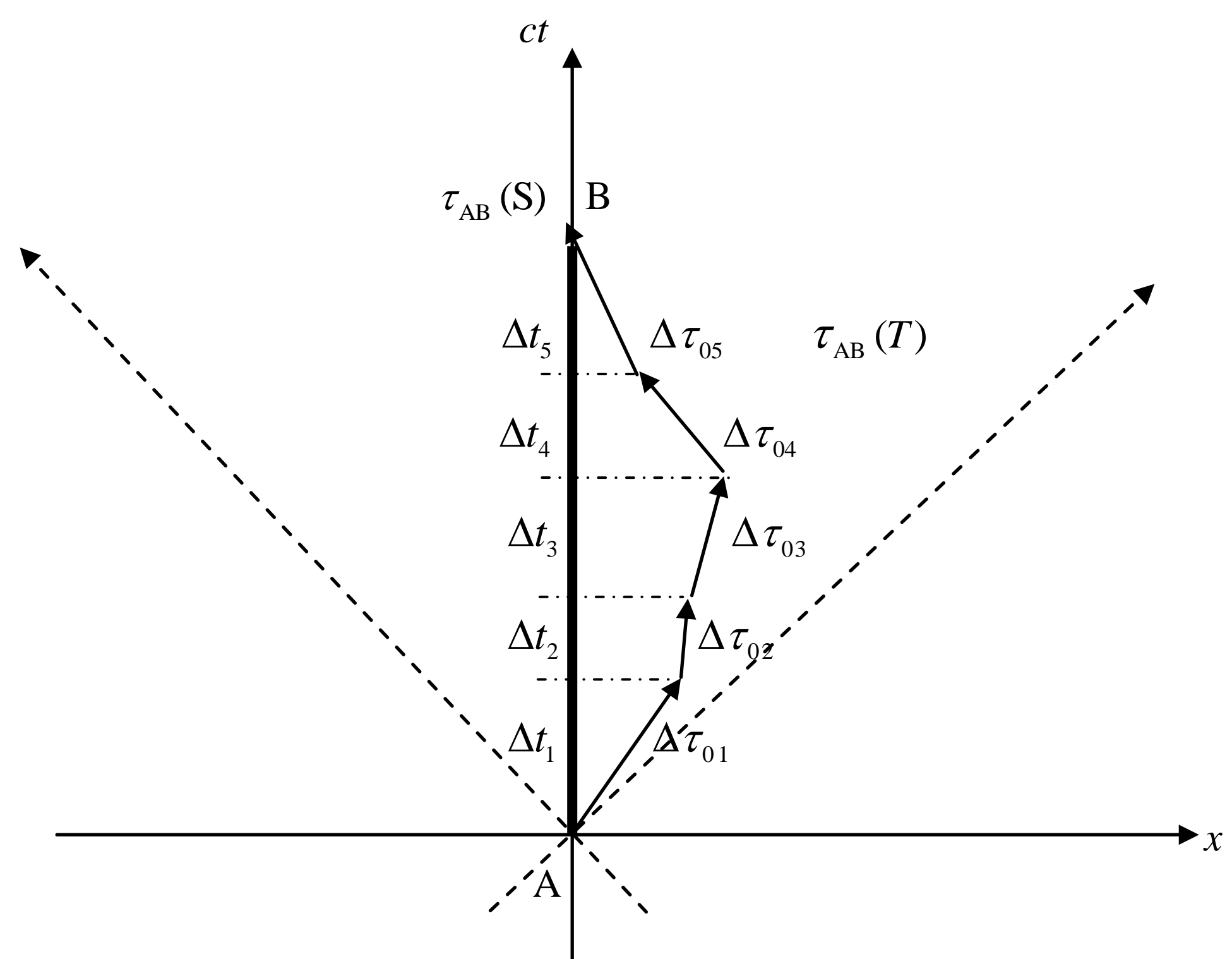

**Fig. 2**

The proper time of an object T moving in an inertial frame $K_0$ in which the initial and final moment of motion (events A and B) occur in the same place (proper frame).

The intermediate stages of motion of T are represented by a zig-zag line connecting A and B. For T returning to its starting point this is just a generalized version of the twin (or clock) paradox with one clock S stationary in $K_0$ and the other one (T) making a round trip. The proper time $\tau_{AB}(\mathrm{S})$ read by S is greater than the net proper time $\tau_{AB}(\mathrm{T})$ of T between A and B. The time $\tau_{AB}(\mathrm{S})$ is the *geometric* (not algebraic!) sum of the proper times $\Delta\tau_{0j}$ of individual inertial stages of T's trip, which equals the *algebraic* sum of their projections (dilated times $\Delta\tau_j$) onto the *ct*-axis. The vectors representing $\Delta\tau_{0j}$ are generally not parallel to the *ct*-axis due to T's motion along $x$ during the corresponding stages. The dashed lines are the world lines of photons passing thorough the origin.

Let $\mathbf{r}$ denote the net *spatial* displacement between A and B. Just as setting the net *momentum* $\mathbf{p} = \mathbf{p}_0 = 0$ for the system of masses (3.11) determines their rest frame, setting $\mathbf{r} = \mathbf{r}_0 = 0$ determines the *proper inertial frame* $K_0$ for *process* AB. The term "proper" here reflects the fact that $K_0$ is the rest frame of object S *remaining at* O during the whole round trip of T. This draws a sharp distinction between frame $K_0$ thus defined and frame K co-moving with T, which is *not* inertial. The *net* 4-displacement $s_{AB}$ of T (the geometric sum of its incremental 4-displacements) is identical to that of S, but their respective proper times are different. One can see from Fig.2 that $s_{AB}$ is equal (up to the constant *c*) to proper time of S between events A and B. In other words, it is the proper time $\tau_{AB}(S)$ along the world line of S. Alternatively, this proper time can be obtained as the sum of *dilated* times of the individual sub-processes in T as observed from $K_0$ :

$$\tau_{AB}(S) = \sum_j \Delta t_j = \sum \Delta\tau_j \; \gamma(v_j) \;\; \geq \;\; \sum_j \Delta\tau_j = \tau_{AB}(T) \qquad (3.12)$$

The algebraic sum of their *proper* times $\Delta\tau_j$ on the right of Eq. (3.12) amounts to the proper time $\tau_{AB}(T)$ along the world line of T.

Eq-s (3.11) and (3.12), while describing quite different physical characteristics, are mathematically identical. The ratio of proper time of a stationary object to that of an object in a round-trip travel is analogous to the ratio of the rest mass of a composite system to the sum of the rest masses of its constituents. Like the rest mass, the proper time is a non-additive characteristic of an arbitrary process as measured from $K_0$. Explicit formulation of this analogy gives another view of and another way for explaining the twin paradox. It may be helpful in demystifying the paradox when teaching it to students. From the viewpoint of (3.12) this "paradox" (*dependence of proper time* $\tau_{AB}$ *on path connecting* A *and* B [1, 3]) is just another manifestation of non-additivity of proper time, similar to non-additivity of the rest mass. In the example illustrated in Fig. 2, the net proper time measured by the S-clock is the proper time of a *Stationary* twin (Sam) residing at O. The *algebraic* sum of the proper times of the incremental sub-processes measured by the T-clock is the proper time of the *Traveling* twin (Tom). Just as the rest mass of a system is greater than the sum of individual rest masses of its moving constituents, the proper time of the stationary twin is greater than the sum of the consecutive proper times (net proper time) of traveling twin between their parting and reunion. Just as the rest mass of a system is exactly equal to the sum of *relativistic* masses of its moving parts, the proper time $\tau_{AB}(S)$ is exactly equal to the sum of *dilated* times of the consecutive incremental sub-processes within moving object T. Relation (3.1) between the relativistic mass of an object and its rest mass is identical to relation (3.10) between the dilated time of a process and its proper time; this is natural consequence of the fact that both – mass and time – are the temporal components of their respective 4-vectors (momentum and displacement).

The above-mentioned trend to discard the concept of relativistic mass might be one of the factors that blocked this beautiful analogy from view even though it was exposed since the onset of the theory of relativity, begging for recognition.

III *c*: **Proper distance**. Consider now two events A and B separated by a *space-like* interval, that is

$$s_{AB}^2 = c^2 t_{AB}^2 - \mathbf{r}_{AB}^2 \; < \; 0 \qquad (3.13)$$

In such case we can find a RF where both events happen simultaneously, and the interval (3.13) reduces to pure distance $r_{\text{AB}}$ between them. In this RF we have $t_{\text{AB}} = 0$ and get

$$r_{\text{AB}} = |s_{\text{AB}}| \tag{3.14}$$

The value $r_{\text{AB}}$ defines the *proper distance* between two events [1, 15]. The corresponding inertial frame $K_0$ where $t_{\text{AB}} = 0$ can be called the *proper* frame.

Combining (3.13) and (3.14) shows that proper distance is the norm of the corresponding 4-vector $s_{\text{AB}} = (ct_{\text{AB}}, \mathbf{r}_{\text{AB}})$:

$$r_{\text{AB}}^2 = \mathbf{r}_{\text{AB}}^2 - c^2 t_{\text{AB}}^2 \tag{3.15}$$

We can write it as

$$r_{\text{AB}}^2 = r_{\text{AB}}^2 \left(1 - \frac{c^2}{u^2}\right) \quad , \quad u = \frac{r_{\text{AB}}}{t_{\text{AB}}} > c \tag{3.16}$$

Here *u* can be considered as the speed of a fictitious superluminal particle (tachyon) connecting the end points of a space-like interval (3.13).

Expression (3.16) seems to differ from (3.7) and (3.10). But this is only an apparent difference since (3.16) is expressed in terms of superluminal speed of a tachyon. We can also express it in terms of the relative speed *V* between frames K and $K_0$. From Lorentz transformation between these frames, one obtains the simple relation between *u* and *V* [2, 5]

$$uV = c^2 \tag{3.17}$$

Using this, we can bring (3.16) to the form

$$r_{\text{AB}}^2 = \frac{r_{\text{AB}}^2}{\gamma^2(V)} \quad ; \qquad r_{AB} = r_{\text{AB}}\, \gamma(V) \tag{3.18}$$

Now we see total mathematical equivalence between (3.15), (3.18) and (3.6-8), (3.10). The proper distance can be either measured directly in the proper frame $K_0$ or computed as the norm of $s_{\text{AB}}$ through (3.15); alternatively, it can be computed from $r_{\text{AB}}$ and *V* using (3.18). Thus, the proper distance is in one camp with the rest mass and proper time. And this analogy also extends onto the situations when the interval (3.15) is the geometric sum of a number of sub-intervals $\Delta s_j$ (Fig. 3). Then we can express the net proper distance $r_{\text{AB}}$ as the sum of the spatial components $\Delta r_j$ of these sub-intervals:

$$r_{\text{AB}} = \sum_j \Delta r_j = \sum_j \Delta r_j\, \gamma(V_j) \neq \sum_j \Delta r_j \tag{3.19}$$

Here $V_j$ is relative velocity between the proper frame $K_0$ of $s_{AB}$ and proper frame $K_j$ of a sub-interval $\Delta s_j$. As seen from Fig.3, the above-considered analogy between the rest mass and the proper time can be extended to the proper distance: the latter, while being an invariant, is not an additive characteristic of a system of events.

Summarizing this part, we can say that each of the rotational invariants – $m_0$, $\tau_{AB}$, $r_{AB}$ – can be computed as the norm of its respective 4-vector from its components. All these three characteristics share with *c* its invariance under 4-rotations. And yet they are fundamentally different from *c*: the latter is the norm of a 4-velocity of *any* object; and by virtue of being single-valued it does not require any computing.

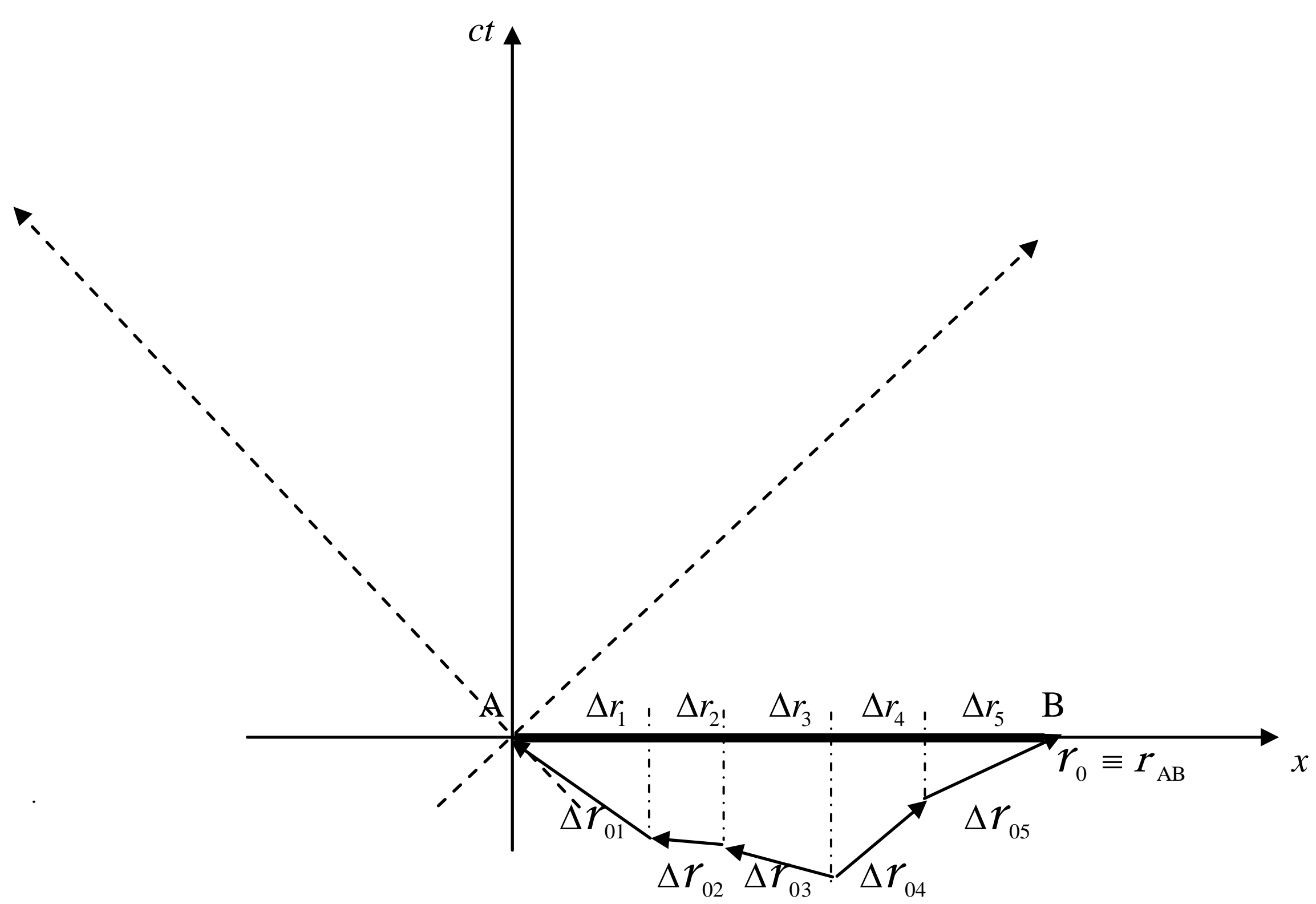

**Fig. 3**

The proper distance between the ends A and B of a space-like 4-displacement.

It is the distance between A and B in the inertial frame $K_0$ (proper frame) in which these events are simultaneous. For a displacement consisting of a number of incremental displacements, the net proper distance $r_0 \equiv r_{AB}$ is the modulus of *geometric* (not algebraic!) sum of these displacements, or, which is the same, the *algebraic* sum of their projection moduli $\Delta r_j$ onto the *x*-axis.

The dashed lines represent space-time trajectories of the photons passing through the origin.

## IV. ***"Face" three* + 1**

Finally, we turn to the fourth category. The quotation mark in its name reflects the fact that it is merely a mask rather than a real face, with no actual invariant under it. The mask covers the *proper length* of an object. This well-known observable has been *proclaimed* an invariant due to a general oversight. The proper length has *nothing to do* with the proper distance described in the previous section. There is a widely spread confusion between these two different concepts. It has reached such a scale that we can describe it by the words of D. Hestenes (said on a different but related occasion [21]) as "the conceptual virus" spread in the Physics community. This virus is manifest in numerous misleading or just wrong statements even in some authoritative sources. Here is one taken from Wikipedia in 2009:

"*In relativistic physics,* ***proper length*** *is an invariant quantity which is the rod distance between spacelike-separated events in a frame of reference in which the events are simultaneous*."

The first part of this statement claims that it is about *proper length*. The second part is the definition of the *proper distance*. The combination of two, as we will see later, makes the whole statement, using Pauli's famous expression, not even wrong. In particular, as the rod's state of motion is not specified here, it may as well be a *moving* rod whose edges are instantly coincident with two events simultaneous in the given frame, in which case the quoted statement defines the proper distance as the *Lorentz-contracted* length of the rod rather than its proper length.

Here are two more quotations, this time from the Forum of Physics Educators [22]:

"…the Lorentz-contracted length is merely a spatial projection of a 4-displacement with its norm being the *proper length*."

"The" length of a ruler is invariant under ordinary rotations, even though the projection of the ruler onto this-or-that reference frame will in general change. In a profoundly analogous way, "the" length of rulers and "the" timing of clocks is invariant under rotations in the XT plane, i.e. boosts, i.e. changes in velocity. It is only the shadow cast on this-or-that reference frame that changes. Do not confuse a shadow with the real object that casts the shadow".

Such statements come from and are shared by a significant part if not majority, of the physics community. Both statements are false, and, when taught to students, promote the above-mentioned viral infection to the status of the world-wide pandemic.

That the first statement is false, is immediately evident from a very simple observation: in Lorentzian geometry, the norm of a space-like 4-displacement is *less* than its spatial projection (see (3.16) or (3.18)), whereas the proper length of a rod is *greater* than its Lorentz-contracted length. Therefore, the proper length is *not* the norm of a 4-vector, and the Lorentz-contracted length is not its "shadow".

For the same reason the second statement, making the proper length "profoundly analogous" to the invariant length under ordinary 3-rotations, is also false. And the claim that *proper length* is in one company with *proper time* makes it double-false.

One of the possible sources of confusion between the two characteristics is the fact that both – Lorentz contraction and time dilation – are consequences of the relativity of time, and each of them implies the other one [2, 5]. This correlation, however, does not imply identical physical behavior, and indeed, we know that proper time, in contrast to proper length, is shorter than its "shadow" – the dilated time. Nevertheless, the *logical* correlation of the two effects was misconstrued by many into *physical* equivalence between them with promoting proper length to the same status as proper time. This automatically led to wrongly identifying the proper length with proper distance, which is, indeed, in one camp (Sec.3C) with proper time.

Probably the best cure against the confusion between these totally different concepts would be reformulation of physics in the language of Geometric Algebra [21, 23-25], but this requires time and concerted effort. Here we will just show that the above misconceptions can be easily clarified within the framework of existing formulation. We consider a few examples illustrating the difference between proper distance and proper length.

Take a rigid rod of proper length $l_0$. If we are in the rest frame $K_0$ of the rod and want to measure its length, it is *not* necessary to mark the end-points simultaneously in $K_0$. Moreover, if we are to consider the $l \leftrightarrow l_0$ (length – proper length) relationship for a fixed 4-displacement with simultaneous marking of the rod's ends in a moving frame K, then Relativity *demands* these markings to be *not* simultaneous in $K_0$.

Thus, we can mark one end now (at a moment $t_1$) and the other end later (at a moment $t_2 > t_1$) in $K_0$, and then measure the distance between the marks. We can even consider such measurement as a possible operational definition of proper length. From the viewpoint of experimental physics, the requirement that the marks be made simultaneously is redundant for a stationary object of constant shape and size, and it can in this case be dropped. The distance between the ends of the stationary rod is its proper length *regardless* of the time lapse between the two markings. But this time lapse, together with the spatial separation $l_0$, determines 4-displacement between the two corresponding events (Fig. 4).

Since $l_0$ is fixed while the time lapse is allowed to be arbitrary in $K_0$, we have an infinite set of possible different 4-displacements with common spatial projection $l_0$. In other words, *the same proper length* can be a spatial projection of an infinite number of *different* 4-vectors with different temporal components and accordingly different norms. By the same token, it is *not* the proper distance between the marking events if the marks are not made simultaneously in $K_0$. Moreover, we can make the time separation between the markings so big, that the corresponding 4-displacement becomes light-like or even time-like! Consider a stationary rigid rod of 1 m proper length. In its rest frame, we mark its left end now and its right end one million years from now. The 4-displacement between the markings is definitely time-like, in which case it cannot even be assigned a proper distance. There is no such thing as a proper distance for a time-like interval! And yet we can measure the spatial distance between the marks and obtain exactly 1 m, which is, according to definition, the proper length of the rod. Thus, the proper length here is neither the proper distance (which does not exist for OC and OD in Fig. 4), nor the norm of the corresponding 4-displacement, which is time-like!

Consider now a reciprocal situation: let the rod move with velocity $V$ along the $x$-axis of a reference frame K and be aligned with the direction of motion. As before, we can measure the rod's length in K by marking its edges and then measuring the distance between the marks; but now, since the rod is moving, it is absolutely imperative that the instant positions of its edges be marked simultaneously in K. In this case, the spatial separation $r_{AB} = l$ between the marks is, by definition, the *proper distance* between the marking events; but it is *not* the proper length of the rod! Indeed, the described procedure constitutes the length measurement of the moving and accordingly *Lorentz-contracted* rod; the rod's *proper* length $l_0$ (assuming its speed is known) is obtained from $l$ as:

$$l_0 = \gamma(V)l = \gamma(V)r_{AB} \neq r_{AB} \tag{4.1}$$

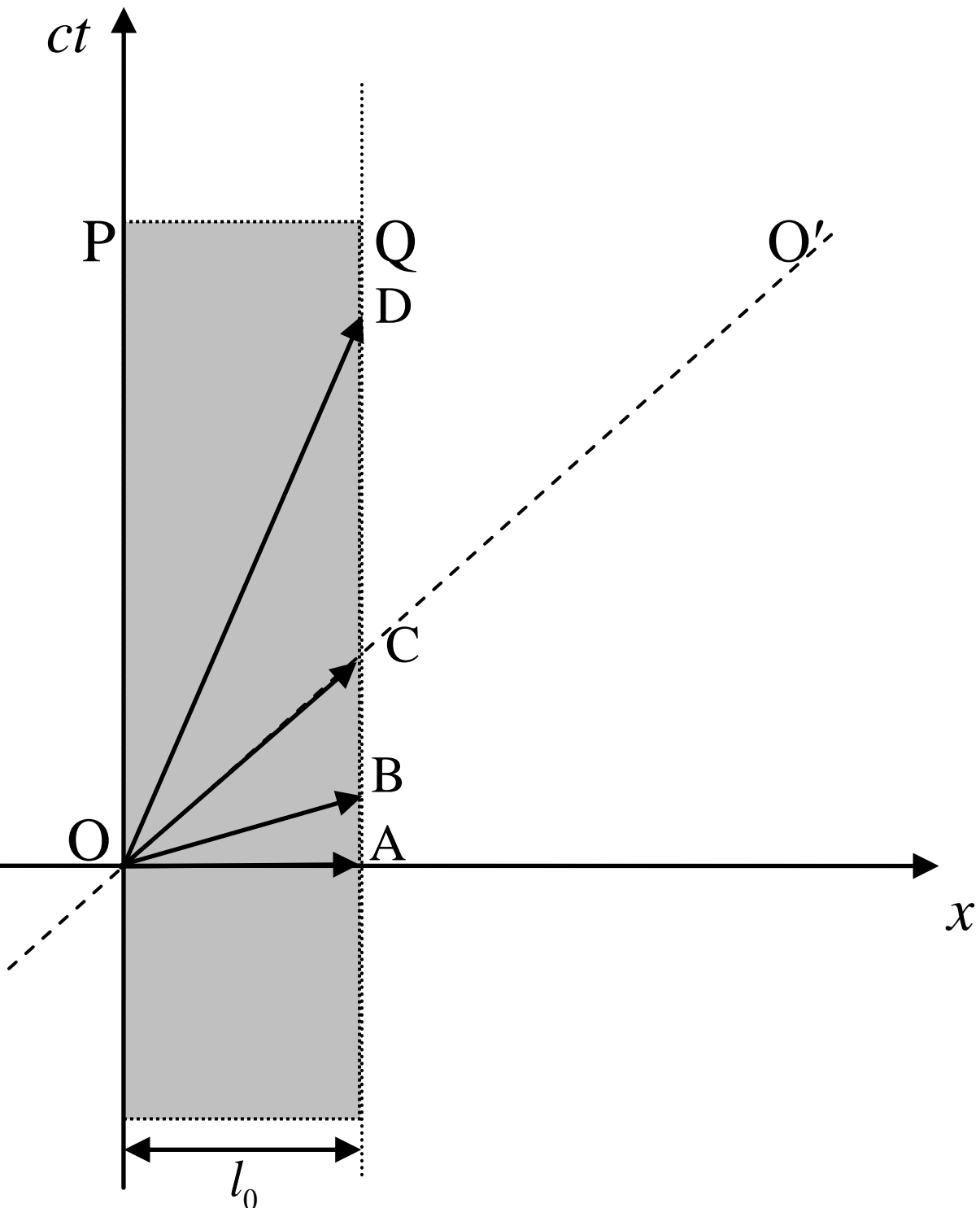

**Fig. 4**

The world sheet of a stationary rod, with a proper length OA = $l_0$ aligned along the $x$-axis. OP and AQ are the world lines of the rod's edges. OO′ is the world line of a photon passing through the origin. The width of the sheet (the proper length $l_0$) is the common spatial projection of space-like 4-displacements OA and OB, light-like 4-displacement OC and time-like 4-displacement OD. In case OA the corresponding 4-displacement is coincident with the proper length. In neither case (except for OA) is the proper length equal to the norm of the corresponding 4-displacement. In particular, the norm of OC is zero. And in neither case, except for OA, is the proper length identical to proper distance even numerically, let alone conceptually. Thus, 4-displacement OC has in the described case the proper length $l_0$ as its spatial projection, but no proper distance is associated with it (there is no RF in which the events O and C would be simultaneous). The same is true about the time-like 4-displacement OD.

Again, the proper length here is not the norm of a 4-vector.

The non-trivial relationship between proper length and proper distance can be seen in more details from the space-time diagram of the moving rod (Fig. 5).

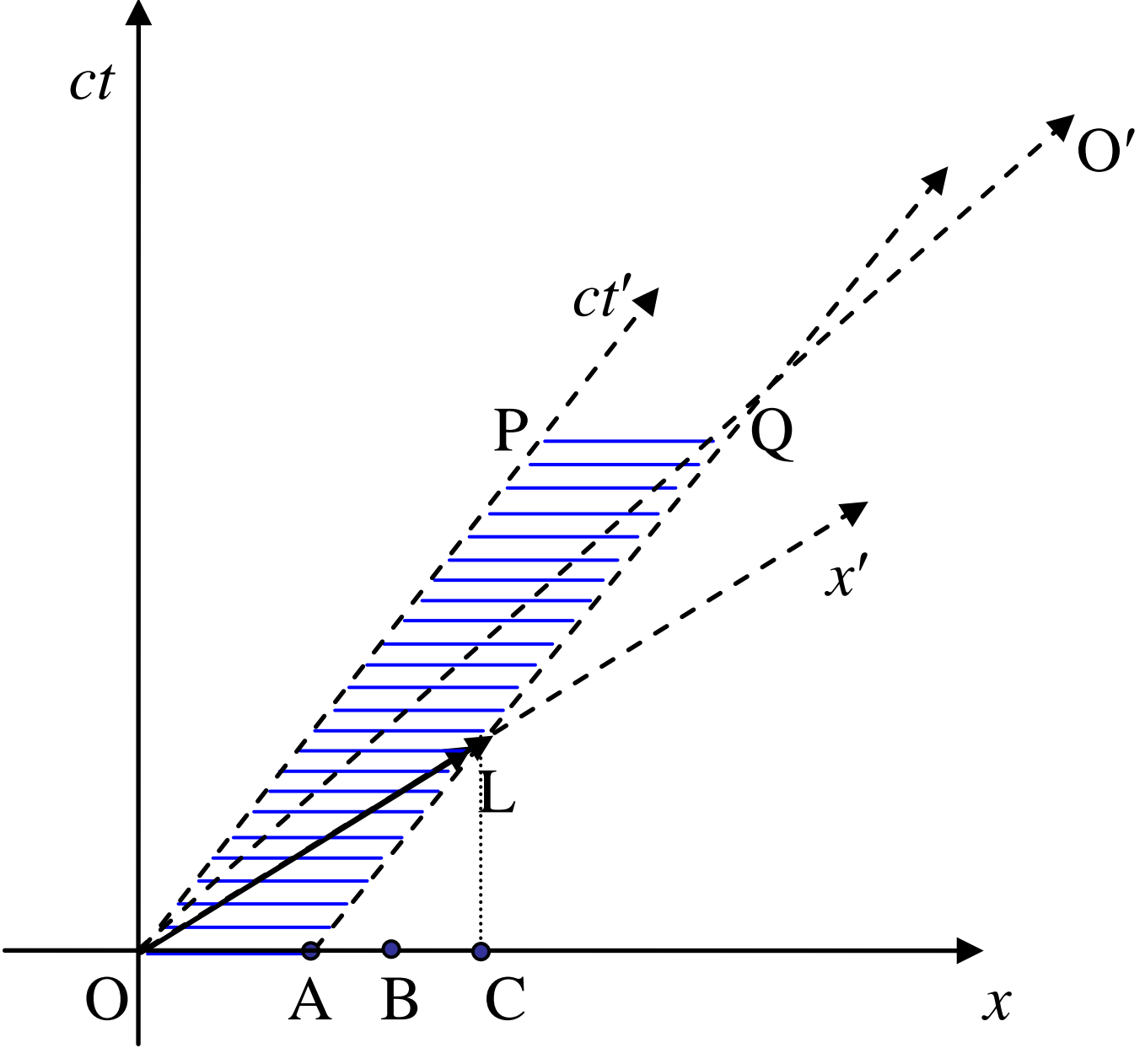

**Fig. 5**

The rod from Fig. 4 as observed from a reference frame K where it is moving along $x$ with a speed $V$. O-$ct'$ and O-$x'$ are the temporal and spatial axes, respectively, of the rest frame $K_0$ of the rod (as observed from K), and OAPQ is its world sheet in K. $\mathrm{OA} = l_0/\gamma(v)$ is the Lorentz-contracted length of the rod. OB represents the proper length of the rod if it were stationary in K. OC is the spatial projection of the 4-displacement OL between two markings of the rod's ends made simultaneously in $K_0$.

The diagram shows the world sheet of such a rod, with the segment

$$\mathrm{OA} = l = l_0/\gamma(V) \equiv x_{\mathrm{A}} \tag{4.2}$$

being its Lorentz-contracted length. The primed axes $ct'$ and $x'$ are now the temporal and spatial axes, respectively, of the rod's rest frame $K_0$. Since events O and L happen at the edges of the rod which is stationary in $K_0$, the norm of 4-displacement OL is equal to its proper length. Since these events are simultaneous in $K_0$, the norm is also equal to the proper distance between O and L. This is one of the few special cases when the proper length is coincident with proper distance. But even in this case the Lorentz-contracted length OA, contrary to the cited statement from [22], is not the spatial projection of OL. Apart from being immediately evident from Fig.5, this can easily be proved quantitatively. Assume as usual that the local clocks at the respective origins of both frames K and $K_0$ read the zero time when they are instantly coincident (event O). Denoting the spatial projection of OL as $\mathrm{OC} \equiv x_{\mathrm{L}}$ and the time coordinate of event L in K as $t_{\mathrm{L}}$, we have

$$x_{\mathrm{L}} = x_{\mathrm{A}} + V t_{\mathrm{L}} \tag{4.3}$$

Lorentz-transformed time-coordinate of this event in $K_0$ is

$$t'_{\mathrm{L}} = \gamma(V)\left( t_{\mathrm{L}} - \frac{V}{c^2} x_{\mathrm{L}} \right) \tag{4.4}$$

Since $x'$ is the line of simultaneity in $K_0$, we have $t'_{\mathrm{L}} = 0$, which gives

$$t_{\mathrm{L}} = \frac{V}{c^2} x_{\mathrm{L}} \tag{4.5}$$

Putting this into ((4.3) and solving for $x_{\mathrm{L}}$ yields, in view of (4.2)

$$x_{\mathrm{L}} = \gamma(V)\, l_0 \tag{4.6}$$

As was already mentioned, not only is $x_{\mathrm{L}}$ greater than $l = x_{\mathrm{A}}$, it is greater than $l_0$.

As a by-product of this derivation we see that the proper length (represented by OB in K) is the geometric mean of the Lorentz-contracted length and the spatial projection of 4-displacement OL. Thus, the spatial projection $r_{12}$ of a proper distance $\mathcal{r}_{12}$ is, as it should be in Lorentzian geometry, *greater* than $\mathcal{r}_{12}$, and the Lorentz-contracted length is *not* a projection of the proper length. The latter *cannot* be computed from *l* and *t* in the same way as (3.15).

Strictly speaking, we cannot even claim that the relativistic length contraction is a purely geometric effect. It is rather a combination of geometry and dynamics, which is especially evident in length measurements of accelerated objects or objects of varying size [5, 26, 27].

Thus, the *proper distance and the proper length describe quite different characteristics* of a process or an object. The former relates to a pair of events in space-time, which are connected by a space-like interval; the latter describes geometrical properties of a material object observed in its rest frame. It is not an invariant of category III (let alone I or II !). It could be named the "conditional invariant" merely by convention: when we measure the Lorentz-contracted length *l* of a moving object, we just remember in the back of our mind that the observer sitting on this object would record the length $l_0$. In other words, we mentally substitute the actual length *l* of the object in a given RF by its length $l_0$ in the co-moving RF. If we know *V*, we can compute $l_0$ from *l* using (4.2), but this computation has nothing to do with those in Part III. If we apply consistently the logics of described "promotion" (naming the "invariant" any characteristic of an object in its rest frame) then nearly all physical quantities will become "invariants"; in which case the mere concept of invariance will be stripped of its meaning. But once the proper length has been universally acclaimed as an invariant, we must at least be very careful to separate it from true invariants of type I, II, and III, and put it into a separate category. Naming things as they are, we should call this type the *bogus* invariant. More politely (until the proper length is disqualified from its current status) we can call it *conventional* (or *conditional*) *invariant*.

## Conclusion

Three distinct types of Lorentz-invariance under 4-rotations, plus one included into this family by collective mistake, are summarized in the tables 1 and 2 below:

**Table 1: True invariants**

| | Invariant Type | Physical characteristic | Computational algorithm |
|---|---|---|---|
| I | Operational (frame-independent by measurement) | Number of particles $N$<br>Electric charge $Q$ | Unknown or not identified |
| II | Absolute (universal) | Speed of light $c$ | Satisfies the identity: $c^2 \equiv u_0^2 - \mathbf{u}^2 = u_j u^j$ |
| III | Rotational (directly measurable in the rest (proper) frame and computable as the norm of a 4-vector in an arbitrary inertial RF) | Rest mass $m_0$<br>Proper time $\tau_0$<br>Proper distance $r_0$ | $m_0^2 = m^2 - \left(\mathbf{p}/c\right)^2$ or $m_0 = m/\gamma(v) \le m$<br>$\tau_0^2 = t^2 - \left(\mathbf{r}/c\right)^2$ or $\tau_0 = t/\gamma(v) \le t$<br>$r_0^2 = \mathbf{r}^2 - \left(ct\right)^2$ or $r_0 = r/\gamma(v) \le r$ |

**Table II: False invariants**

| Invariant Type | Physical characteristic | Computational algorithm |
|---|---|---|
| Bogus (does *not* satisfy the definition of an invariant) | Proper length $l_0$ (must be expelled from the invariants' family) | $l_0^2 \neq \mathbf{r}^2 - (ct)^2$**!** $l_0 = l\,\gamma(v) \ge l$ |

*Summary*: Operational invariants (I) are directly measurable and remain the same in *any* RF. Absolute invariant $c$ (II) is directly measurable for an object in motion if $m_0 = 0$ or *computable* as the norm of its 4-velocity in *any* state of motion (including rest) if $m_0 \neq 0$. Rotational or norm-invariants (III) are directly measurable in a proper or rest frame of a system and computable as the norm of the corresponding 4-vector characterizing the system.

## References

1. E. Taylor, J. A. Wheeler, *Exploring Black Holes*: *Introduction to General Relativity*, Addison Wesley Longman, San Francisco (2000), p. 1-13.
2. M. Fayngold, *Special Relativity and Motions Faster than Light*, Wiley-VCH (2002)
3. L. Landau and E. Lifshits, *The Classical Theory of Fields*, Butterworth-Heinemann (1998)
4. P. G. Bergmann, *Introduction to the Theory of Relativity*, Dover Publications, Ch. 6 (1974)
5. M. Fayngold, *Special Relativity and How It Works*, Wiley-VCH, Weinheim (2008)
6. T. Sandin, In Defense of Relativistic Mass, *Am. J. Phys*., **59** (11), 1032 – 1036 (1991)
7. C. Moeller, *The Theory of Relativity*, Oxford Univ. Press (1962), Ch. 6
8. R. C. Tolman, *Relativity*, *Thermodynamics*, *and Cosmology*, Clarendon Press, Oxford (1962)
9. M. A. Tonnelat, *The Basis of Electromagnetism and Relativity Theory* (1962)
10. L. Landau, E. Lifshits, *Hydrodynamics*, Butterworth-Heinemann (1999), Ch. 15, Sec. 133
11. K. Kaufmann, Die Electromagnetische Masse des Electrons., *Phys*. *Zeitschr*. **4**, p. 54 (1902)
12. G. Neumann, Die trage Masse schnell bewegter Electronen, *Ann*. *Phys*. **45**, p. 529 (1914)
13. E. M. McMillan, The Synchrotron – A Proposed High Energy Particle Accelerator, *Phys*. *Rev*. **68** (5 – 6), 143 – 144 (1945)
14. V. Veksler, Concerning Some New Methods of Acceleration of Relativistic Particles, *Phys*. *Rev*. **69**, p. 244 (1946)
15. E. F. Taylor, J. A. Wheeler, *Spacetime Physics*, Freeman & Co. (1963), Sec. 13
16. G. Oas, On the abuse and use of relativistic mass, arXiv: *Physics*/0504110v2 (2005)
17. L. B. Okun, The Virus of Relativistic Mass in the Year of Physics. In the "*Gribov Memorial Volume*: *Quarks*, *Hadrons*, *and Strong Interactions*", World Scientific, Singapore (2006)
18. R. Feynman, *The Feynman Lectures on Physics*, Vol. I, Addison – Wesley (1963), Ch. 16, Sec. 5
19. V. A. Fock, *Theory of Space, Time, and Gravitation*, Pergamon Press – Macmillan Co., 2-nd Ed. (1964)
20. F. Wilczek, Mass Without Mass, *Phys. Today*, Nov. 1999, p.11; *ibid*. Jan. 2000, p.14
21. D. Hestenes, Oersted Medal Lecture 2002: Reforming the mathematical language of physics, *Am. J. Phys*. **71** (2), 104 – 121 (2003)
22. "*Teaching Special Relativity*", Forum for Physics Educators, Phys – 1, phys-1@carnot.physics.buffalo.edu , (July 30, 2009)
23. D. Hestenes, Spacetime physics with geometric algebra, *Am. J. Phys*., **71** (9), 691–714 (2003)
24. D. Hestenes, *Space-Time Algebra*, Gordon and Breach (1966)
25. A. Lasenby and C. Doran, *Geometric Algebra for Physicists*, Cambridge Univ. Press (2003)
26. M. Fayngold, Dynamics of relativistic length contraction and the Ehrenfest paradox, arXiv: 0712.3891v1
27. M. Fayngold, Two permanently congruent rods may have different proper lengths, arXiv: 0807.0881 (July 2008)